\documentclass[
 amsmath,amssymb,
 aps,
pre,
twocolumn,
longbibliography,
10pt,a4paper,sort&compress,showpacs,superscriptaddress
]{revtex4-2}
\usepackage{graphicx} 
\usepackage{amsmath}
\usepackage{amssymb}
\usepackage{amsfonts}
\usepackage{bm}
\usepackage{dcolumn}

\usepackage{amsmath,amssymb,amsthm}
\usepackage{mathrsfs}
\usepackage{commath}
\usepackage{dcolumn}
\usepackage{hyperref}
\usepackage{etoolbox} 
\usepackage{siunitx}
\usepackage{multirow}

\usepackage{color}
\usepackage{xcolor}
\usepackage{overpic}

\usepackage[english]{babel}
\usepackage[caption=false]{subfig}
\addto\captionsenglish{}
\usepackage[capitalise,nameinlink]{cleveref}
\newcommand{\aref}[1]{\hyperref[#1]{Appendix}}
\usepackage{blkarray}

\begin{document}
\title{
Data-driven geometric phase in biological locomotion
}

\author{Pyae Hein Htet}
 \affiliation{
	Department of Mathematics, Kyoto University, Kyoto 606-8502, Japan}
\author{Kenta Ishimoto}
\email{kenta.ishimoto@math.kyoto-u.ac.jp}
 \affiliation{
	Department of Mathematics, Kyoto University, Kyoto 606-8502, Japan}
 
\begin{abstract}

Geometric phase quantifies net locomotion in dissipative media via gauge theory, but linking this theoretical quantity to noisy, sparse, and weakly periodic biological shape data is challenging. We develop a theory-guided, data-driven Koopman autoencoder to recover the limit cycle embedded in imperfect cyclic data and extract shape gaits and geometric phase from sperm and nematode data. We introduce a geometric phase sensitivity function that quantifies responses to shape perturbations and reveals mechanical information using only gauge-theoretic structure, without assuming mechanical laws.

\end{abstract}
\date{\today}

\maketitle

\section{Introduction}
Geometric phase, which encodes the transformation history of a physical system, has become an influential concept across a broad range of modern physics~\cite{cohen_geometric_2019}, extending well beyond its original quantum-mechanical context, with examples in acoustics~\cite{ma_topological_2019, xue_topological_2022}, robotics~\cite{kelly_geometric_1995, aguilar_review_2016, li_robotic_2022}, geophysics~\cite{delplace_topological_2017}, and nonequilibrium thermodynamics~\cite{fang_nonequilibrium_2019, wang_geometric_2022}. The hydrodynamics of swimming microorganisms, formulated in terms of gauge theory, has also offered a playground for geometric phase, which in this context corresponds to displacement and rotation over one period of deformation~\cite{shapere_geometry_1989,  hatton_nonconservativity_2015, koens_geometric_2021, ishimoto_bendingcompression_2025}. This mathematical structure arises from rapid viscous dissipation in low-Reynolds-number flow, where the swimming dynamics are solely determined by the swimmer's shape gait. More recently, experimental studies~\cite{dorgan_meandering_2013, keaveny_predicting_2017, zhao_walking_2022, rieser_geometric_2024} have found that the gauge-theoretic structure ubiquitously holds in the class of kinematic locomotion, which encompasses a wider range of biological locomotion in dissipative media, including snakes crawling on sand, insects walking on the ground, and worms burrowing in mud.

With models of body-environment interaction such as the Stokes equation of low-Reynolds-number flow and empirical resistive force theory, one may theoretically compute locomotion dynamics, and thus geometric phase, from a period of shape gait~\cite{lauga_fluid_2020}. Real biological data, however, contain internal and environmental fluctuations and often suffer from limited quality with low spatial and temporal resolutions.
Linear dimensionality reduction via principal component analysis (PCA) has successfully represented shape gait data in (typically two-dimensional) PCA space across systems ranging from cells to  animals and artificial robots~\cite{stephens_dimensionality_2008, werner_shape_2014,  nishiguchi_flagellar_2018, walker_computer-assisted_2020, rieser_geometric_2024}, while finer details of shape information are often captured only by higher PCA modes~\cite{gholami_waveform_2022, ahmad_bio-hybrid_2022, gutierrez_quantification_2025, ishimoto_coarse-graining_2017}. 

In extracting periodic gait as a limit cycle embedded in noisy data, linear methods such as PCA only provide protophase and do not define a proper phase, termed the isochron phase or asymptotic phase~\cite{kralemann_phase_2008}. Hence, nonlinear methods are essential for theoretical consistency. 
In the last decade, the method of phase reduction has been reformulated via Koopman operator theory, which reframes a nonlinear system as an infinite-dimensional linear system and is widely used in data-driven modeling of dynamics~\cite{mauroy_koopman_2020, mezic_koopman_2021, brunton_modern_2022}.
The isochron phase of a limit cycle, in this context, can be obtained via the principal Koopman eigenfunction~ \cite{mauroy_isostables_2013, wilson_isostable_2016, shirasaka_phase-amplitude_2017}. Although a useful and fundamental theoretical construct, Koopman eigenfunctions are notoriously difficult to determine in practice, spurring the development of machine learning approaches.
Recently, a deep-learning architecture has been proposed to construct the isochron phase~\cite{yawata_phase_2024}, and it has also succeeded in extracting limit cycles in fluid and reaction-diffusion systems with large dimensions~\cite{hiruta_autoencoder_2025, yawata_phase_2025}. However, machine learning models usually require large, clean datasets, posing a challenge for data-driven approaches to small-sized biological data. 

In this paper, we develop a theoretically-grounded, data-driven approach to the geometric phase of biological locomotion, using real data from slender locomotors, including sperm cells and crawling nematodes, as prototypical small-sized, low-quality shape data only spanning several beat cycles and exhibiting fluctuations in shape and period. 
After learning the embedded periodic gait from noisy data, we extract the geometric phase via the learned isochron phase, imposing only the gauge-theoretic structure, without assuming any mechanical relations such as fluid equations or a resistive-force-type empirical theory.
Further, as a natural extension of phase reduction theory for geometric phase, we introduce the {\it geometric phase sensitivity function}, which measures the impact of shape perturbations on locomotion. We show that this new physical quantity characterizes the underlying mechanics directly from the data and is useful in understanding the locomotor's steering and maneuverability.

{
The paper is organized as follows. We  introduce the background theory and methodologies in Sec.~\ref{sec:theor}. After presenting the results on synthetic waveform data in Sec.~\ref{sec:synth}, we analyse experimental waveform data in Sec.~\ref{sec:expr}, where we discuss the geometric phase sensitivity function. Sec.~\ref{sec:conc} concludes the study.
}

\begin{figure*}[!tb]
    \centering
    \includegraphics[width=0.8\linewidth]{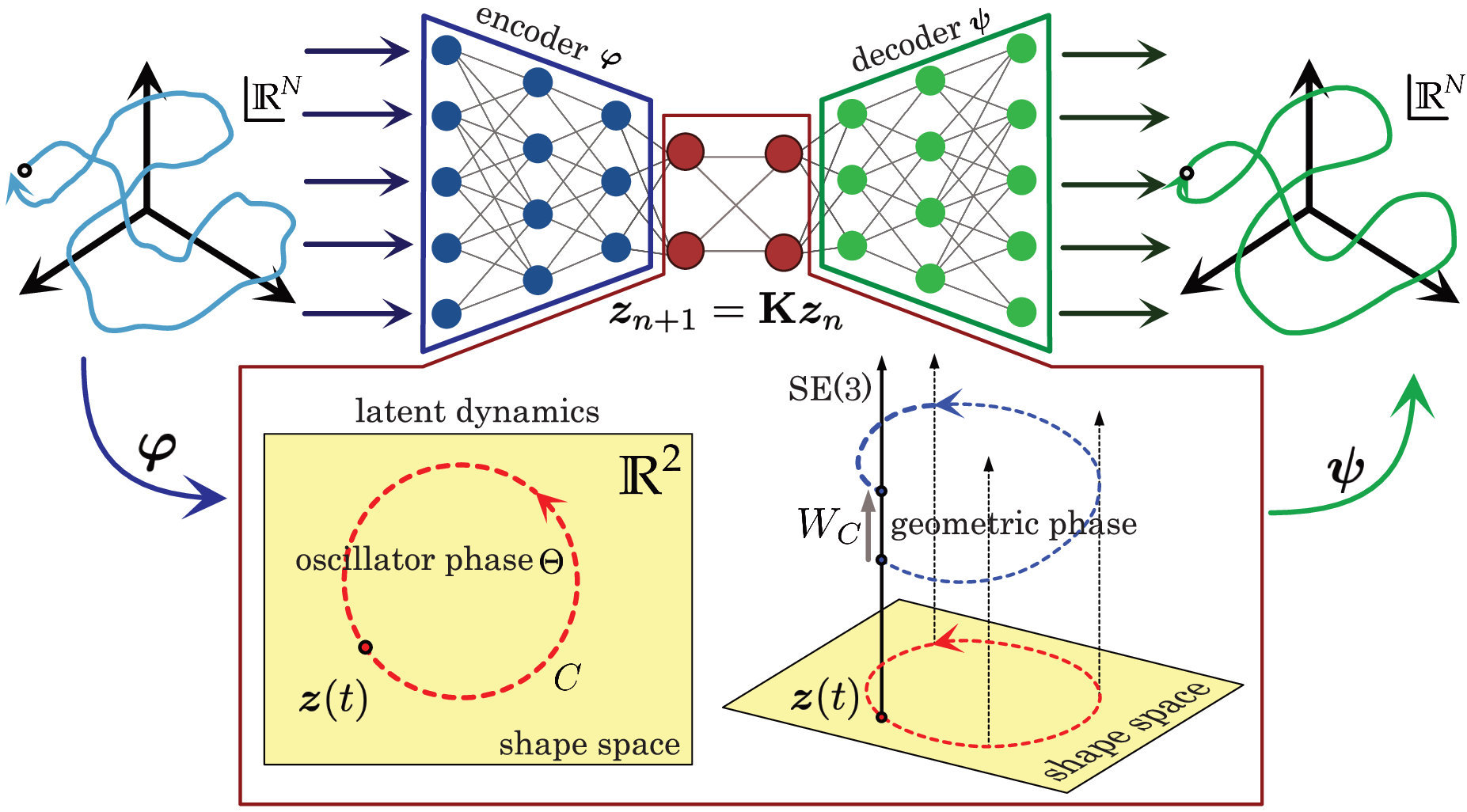}
    \caption{Schematic of theoretical method and numerical architecture. We extract a stable limit cycle embedded in large dimensional shape data, $\bm{q}\in\mathbb{R}^N$, via the principle Koopman eigenfunction $\bm{z}=\bm{\varphi}(\bm{q})\in\mathbb{R}^2$, which maps the shape data into two dimensional latent space, where the time evolution follows linear dynamics and oscillator phase $\Theta$ is defined. The geometric phase $W_C$ is defined as the spatial and orientational shift after one period of deformation. The principal Koopman eigenfunction is numerically constructed by a deep autoencoder, where the map $\bm{\varphi}$ is the encoder and the decoder $\bm{\psi}(\bm{z})\in\mathbb{R}^N$ provides the denoised orbit.}
    \label{fig:schem}
\end{figure*}
    
\section{Theory and methods}
\label{sec:theor}
\subsection{Geometric phase in biolocomotion}

The motion of a self-propelled locomotor is described by its position $\bm{x}(t)\in\mathbb{R}^3$ and orientation ${\bf R}(t)\in {\rm SO}(3)$, represented by $g=({\bf R}, \bm{x})\in {\rm SE}(3)$. 
The shape gait of the locomotor is parameterized by $\bm{q}(t)\in \mathbb{R}^N$, where $N$ denotes the number of shape parameters.
In kinematic locomotion, as in microswimmer hydrodynamics, the motion of a locomotor is  proportional to the deformation velocity. Hence, the linear and angular velocities in the body-fixed frame are provided by
    \begin{equation}
        g^{-1}\dot{g}=A(\bm{q})\dot{\bm{q}}
        \label{eq:recon}.
    \end{equation}
Here, the dot symbol denotes the time derivative and $A(\bm{q})$ is determined only by the instantaneous shape and is mathematically equivalent to a gauge field~\cite{shapere_geometry_1989}, or so-called mechanical connection, which encodes the physics of body-environment interactions.

The shape gait of a locomotor typically consists of the repetition of a time-periodic shape deformation, corresponding to a stable limit cycle $C$ as $\bm{q}(t)\in \mathbb{R}^N$. The displacement and rotation after one deformation cycle correspond to the {\it geometric phase} of the loop $C$, and is obtained by integrating Eq.~\eqref{eq:recon} over the loop as
    \begin{equation}
        W_C =\overline{{\rm P}}\exp\left[ \oint_C A(\bm{q})\,d\bm{q}\right]
        \label{eq:geom_phase}.
    \end{equation}
Here, the symbol $\overline{\rm P}$ denotes the anti-path-ordering operator. The geometric phase $W_C\in {\rm SE}(3)$ emerges from an adiabatic change of shape in dissipative systems, illustrated in Fig.~\ref{fig:schem} as a mismatch of the blue trajectory in the vertical direction after one cycle of the deformation in the shape space. The geometric phase $W_C$ contains translation and rotation and is not a scalar value but represented by a $4\times 4$ matrix~\cite{shapere_geometry_1989}.

\subsection{Phase reduction and Koopman autoencoder}

The essential shape dynamics is therefore represented by a circular orbit in a low-dimensional manifold, and Koopman operator theory guarantees \cite{mauroy_isostables_2013, shirasaka_phase_2017} that 
for a stable limit cycle the principal Koopman eigenfunction constructs a change of coordinates, $\bm{z}=\bm{\varphi}(\bm{q})\in \mathbb{R}^2$, that follows 
$\dot{z}=i\omega z$ with $z=z_1+iz_2\in \mathbb{C}$, where $\omega$ is the frequency of the periodic gait. 
The phase of the limit cycle oscillator, $\Theta(t)$, is introduced as $\dot{\Theta}=\omega$ on the loop $C$, and we call this the {\it oscillator phase} to distinguish from the geometric phase. Oscillator phase can be extended to the entire attracting region of the limit cycle via the isochron function, defined as 
$\Theta(\bm{z})=\arg(z_1+iz_2)$.

Biological locomotion is often contaminated by both internal activity noise and external noise from the environment and the observation process. To approximate the principal Koopman eigenfunction $\bm{\varphi}$ that extracts the limit cycle orbit embedded in noisy biological data, we set up a deep autoencoder neural network that enforces linear dynamics in latent space \cite{takeishi_learning_2017, lusch_deep_2018, li_learning_2019, nayak_temporally_2025}, known as a Koopman autoencoder (KAE). As schematically illustrated in Fig.~\ref{fig:schem}, the encoder $\bm{z}=\bm{\varphi}(\bm{q})$ is designed to approximate the Koopman eigenfunction, which follows linear dynamics in latent space. The decoder $\bm{\psi}(\bm{z})$ provides a two-dimensional manifold embedded in the shape space, functioning as a denoiser for the stochastic time series. 

Given the shape data time series $\bm{q}_n=\bm{q}(t_n)$, we enforce linear dynamics $\bm{z}_{n+1}={\bf K}\bm{z}_n$ for the latent variable $\bm{z}_n=\bm{\varphi}(\bm{q}_n)$ and
train our deep neural network to minimize the total cost function $L_{\rm tot}=L_{\rm rec}+L_{\rm pred}+L_{\rm lin}+L_{\rm rad}$, where
    \begin{eqnarray}
        L_{\rm rec}&=&||\bm{q}(t_n)-\bm{\psi}(\bm{\varphi}(\bm{q}(t_n)))||^2,\\
        L_{\rm pred}&=&||\bm{q}(t_{n+1})-\bm{\psi}({\bf K}\bm{\varphi}(\bm{q}(t_n))) ||^2,\\
        L_{\rm lin}&=&\lambda_{\rm lin}||\bm{q}(t_{n+1})-{\bf K}\bm{\varphi}(\bm{q}(t_n)) ||^2,\\
        L_{\rm rad}&=&\lambda_{\rm rad}||\bm{\varphi}(\bm{q}(t_n))-1 ||^2
    \end{eqnarray}
are reconstruction loss, prediction loss, linearity constraint, and radius constraint, respectively. $\lambda_{\rm lin}$ and $\lambda_{\rm rad}$ are regularization coefficients, and $||\bullet||$ denotes the $\ell_2$ norm. 
We set 
\begin{equation}
{\bf K}=\begin{pmatrix} \gamma \Delta t & -\omega \Delta t \\ \omega  \Delta t& \gamma \Delta t\end{pmatrix},    
\end{equation}
where $\Delta t$
is the time discretization and $\gamma > 0$ accommodates the Floquet contraction of stochastic trajectories onto the limit cycle~\cite{kato_asymptotic_2021}. 
Further details of the network architecture and learning implementations are provided in App.~\ref{app:learning}

\begin{figure*}[!t]
     \centering
     \begin{overpic}
    [width=0.9\textwidth]{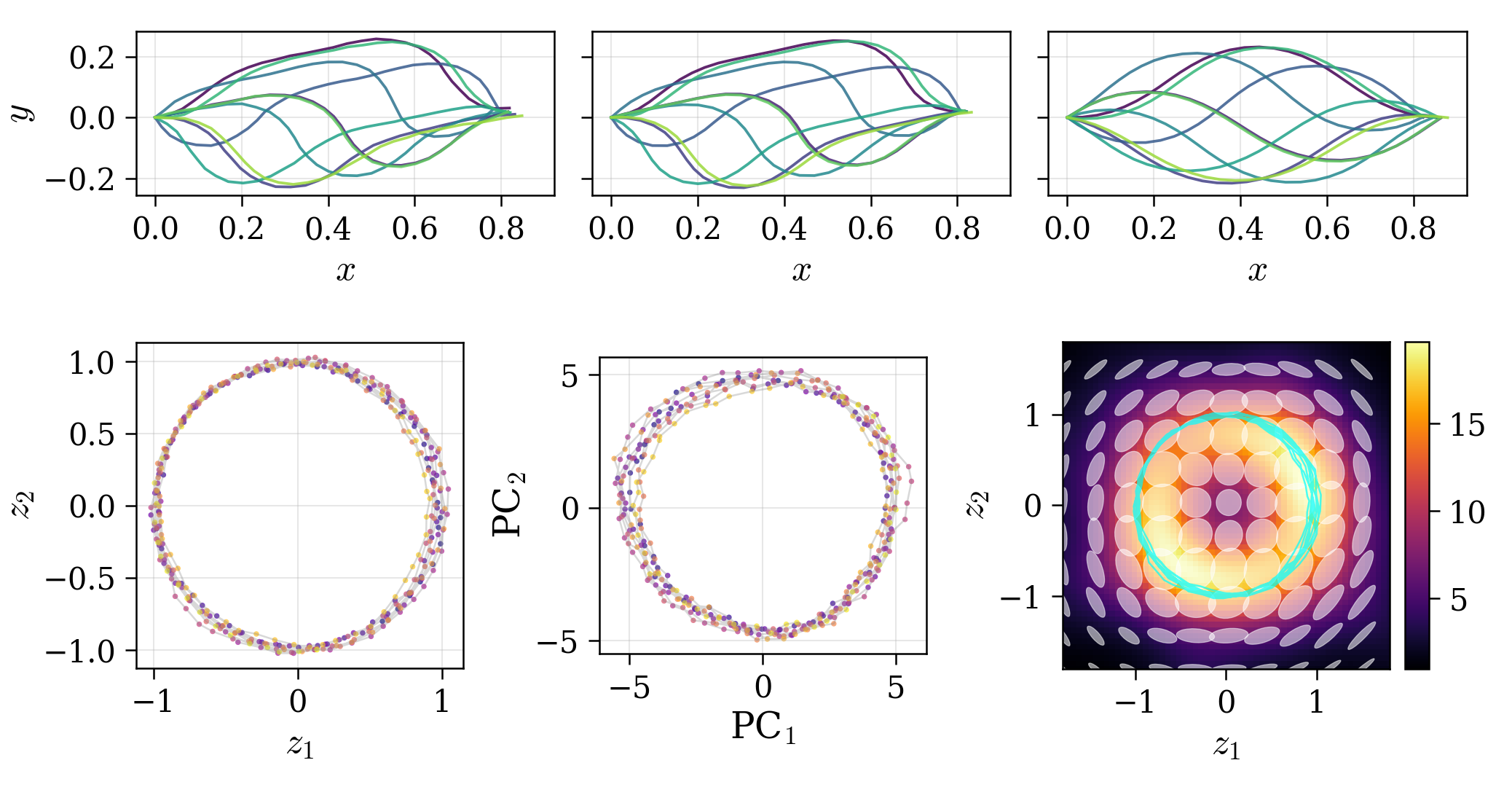} 
    \put(16,51){{\scriptsize original data}}
    \put(49,51){{\scriptsize KAE}}
    \put(80,51){{\scriptsize PCA}  } 
    \put(16,30){{\scriptsize KAE}  } 
    \put(48,29){{\scriptsize PCA}  }
    \put(76,30){{\scriptsize $\sqrt{{\rm det}{\bf G}}$}  }    
    \put(10,51){(a)}
    \put(43,51){(b)}
    \put(75,51){(c)}
    \put(-1,28){(d)}
    \put(32,28){(e)}
    \put(64,28){(f)}    
    \end{overpic}  
     \caption{(a-c) Snapshots of the synthesized waveform, reconstruction by KAE and two-dimensional PCA. (d,e) Data points projected onto the two-dimensional latent space, and two-dimensional PCA space.  (f) Visualization of pull-back metric by $\sqrt{{\rm det}{\bf G}}$ and metric ellipse. Parameters of the synthesized data are set $A_1=0.8$, $A_2=0.4$, $\delta=\pi/2$, $c=1.0$, $b=0.5$, $D_r=0.1$, $D_\phi$=0.1, $D_{\rm obs}=0.05$, $L=1$.}
     \label{fig:synth_mfd}
 \end{figure*}
 
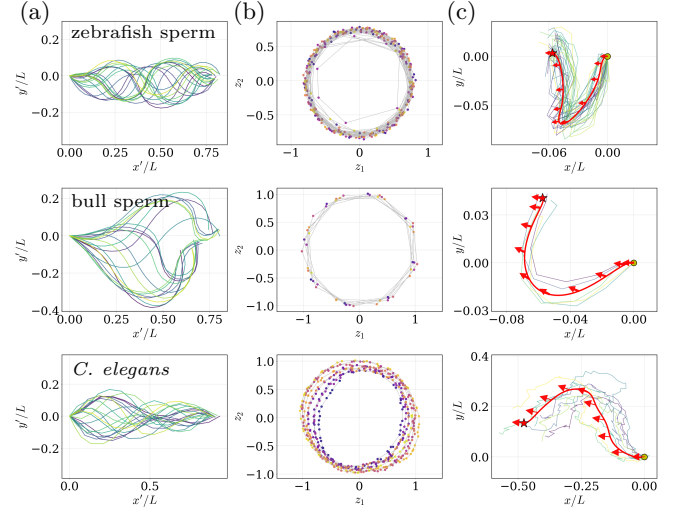
\begin{figure*}[!t]
    \centering
    \begin{overpic}
    [width=0.9\linewidth]{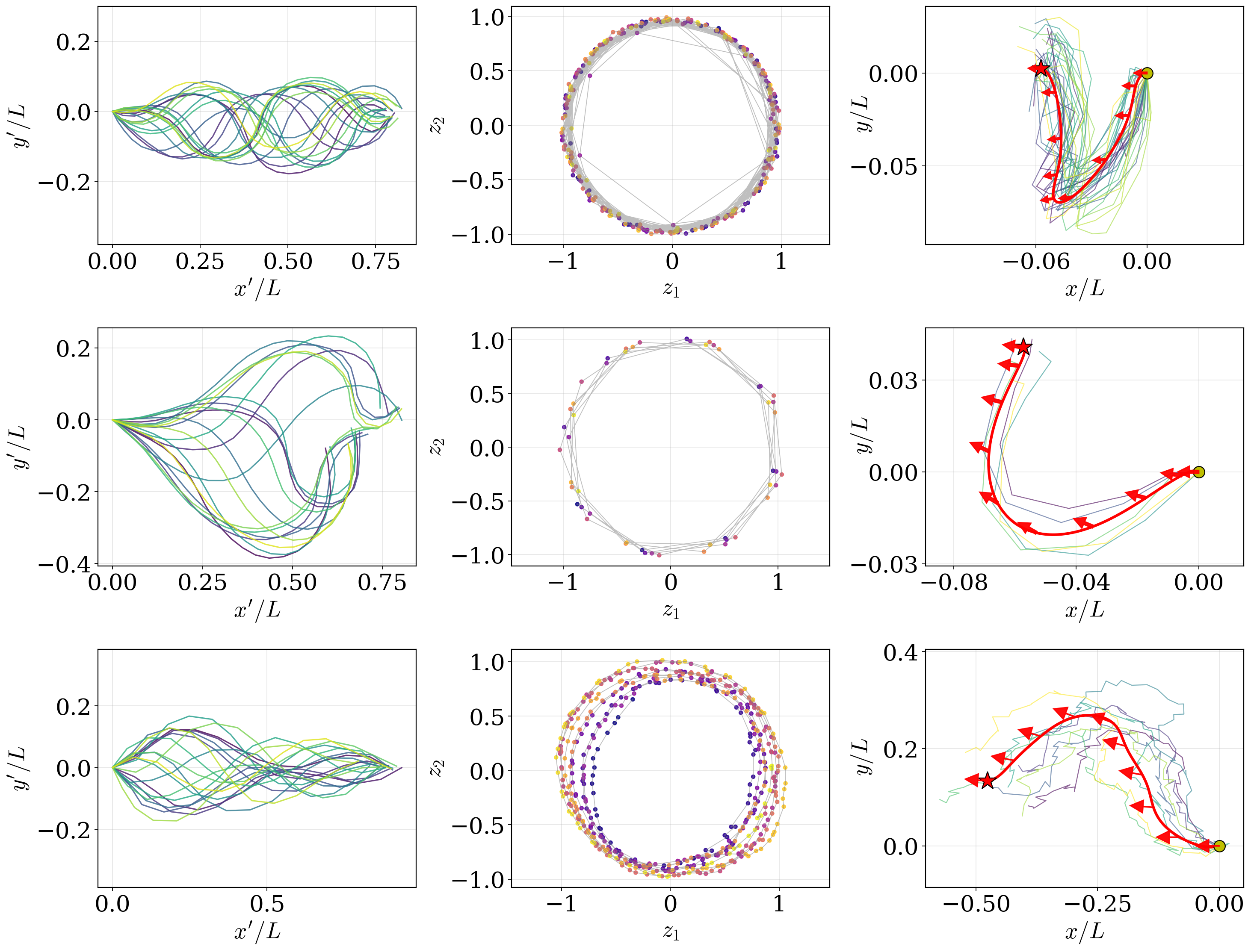}
    \put(9,73){{\scriptsize zebrafish sperm}}
    \put(9,47){{\scriptsize bull sperm}}
    \put(9,21){{\scriptsize {\it C. elegans}}  }  
    \put(1,76){(a)}
    \put(34,76){(b)}
    \put(67,76){(c)}    
    \end{overpic} 
    \caption{ (a) Shape plots in body-fixed frame, (b) scatter plots in Koopman latent space, (c) superimposed trajectories with averaged trajectory obtained by the current method, for zebrafish sperm (top), bull sperm (middle) and {\it C. elegans} (bottom). }
    \label{fig:data}
\end{figure*}
 
\section{Learning synthetic waveform data}
\label{sec:synth}

To validate our methodology, we start with synthetic waveform data of a slender object confined in a two-dimensional space. We let $s\in[0, L]$ be the arclength of the slender locomotor of total length $L$, and take the local tangent angle $q(s,t)$  as our shape variable. Following observations on sperm cells \cite{ma_active_2014}, we assume that the waveform $q(s,t)$ is generated by a noisy limit cycle represented as a normal form of the supercritical Hopf bifurcation, written in polar coordinates $(r,\phi)$ as
    \begin{eqnarray}
    dr=\dot{r}_{\rm det}dt+rD_rdW_r~,~~
        d\phi=\dot{\phi}_{\rm det}dt+D_\phi dW_{\phi}
        \label{eq:synth},
    \end{eqnarray}
with deterministic velocity parts, \begin{equation}
    \dot{r}_{\rm det}=cr(1-r^2) ~,~~ \dot{\phi}_{\rm det}=\omega+b r^2.
\end{equation}
Here, $W_r$ and $W_\phi$ denote the Wiener process, and $D_r$ and $D_\theta$ are the noise intensities.
We then synthesize the tangent angle from this oscillator, with $\zeta=\phi-ks$, as 
    \begin{eqnarray}
        q(s,t)=rA_1\sin(\zeta)+r^2A_2\sin(2\zeta+\delta)+\xi(t),
    \end{eqnarray}
where $k=2\pi/L$ is the wavenumber and $\xi(t)$ is Gaussian noise from observation with noise strength $D_{\rm obs}$. It is readily found by Fourier decomposition that this waveform contains four PCA modes, $\sin(ks), \cos(ks), \sin(2ks),$ and $ \cos(2ks)$.

In Fig.~\ref{fig:synth_mfd}(a), we show snapshots of the synthesized waveform $q(s,t)$, which is spatially discretized by $30$ segments ($N = 29$) and contains approximately 6 beat cycles with $40$ time points per cycle. We also illustrate the waveforms reconstructed by the KAE and two-dimensional PCA [Fig.~\ref{fig:synth_mfd}(b,c)], demonstrating that KAE successfully reproduces the finer details of the waveforms using two latent variables. Further details on  validation of the KAE computations are discussed in App.~\ref{app:valid}

As shown in Fig.~\ref{fig:synth_mfd}(d), 
the trajectory in latent space accumulates near a unit circle, whereas the PCA space trajectory is noticeably more scattered.
Using the Jacobian of the decoder $\bm{\psi}(\bm{z})$, ${\bf J}=\partial \bm{\psi}/\partial \bm{z}$, we can introduce a $2\times 2$ pull-back Riemannian metric ${\bf G}(\bm{z})={\bf J}^{\rm T}{\bf J}$ on the latent space \cite{arvanitidis_latent_2018,shao_riemannian_2018,lee_geometric_2023}, characterizing the nonlinear transformation in extracting the limit cycle. In Fig.~\ref{fig:synth_mfd}(f), we show in color contour $\sqrt{{\rm det}{\bf G}}$, which represents the local expansion. The higher values, shown as brighter colors, reflect nonlinear folding of the orbit in the shape space to fit onto the unit circle in latent space. The pull-back metric ${\bf G}(z)$ is also visualized by ellipses with semi-axes proportional to the two eigenvalues and oriented along the eigenvectors, showing non-uniform embedding in shape space.

\section{Learning experimental waveform data}
\label{sec:expr}
\subsection{Extracting limit cycle in  experimental biological data}

Having demonstrated our method using synthesized data with controlled noise, we now apply our method to extract geometric phase from biological trajectory data, obtained as a movie file recording the organism's shape and motion. We do not assume any mechanical equations such as the Stokes equations or resitive force theory, imposing only the gauge field structure in Eq.~\ref{eq:recon}.

To illustrate the applicability of our data-driven method, we use sperm flagellar waveform data from zebrafish ({\it Danio rerio}) \cite{guasto_flagellar_2020} and bull ({\it Bos taurus})  \cite{guasto_flagellar_2020, friedrich_high-precision_2010}, and body-shape data from the nematode {\it Caenorhabditis elegans} crawling on agar gel from the Supplementary Movie in Ref.~\cite{pierce-shimomura_genetic_2008} (see App.~\ref{app:dataacq}). Snapshots of the waveforms are shown in Fig.~\ref{fig:data}(a). These data are chosen as representatives of external fertilizers, internal fertilizers, and non-swimmer organisms, respectively. Although zebrafish and bull sperm dynamics are well described by the Stokes equation, we do not use this knowledge {\it a priori}. Also, the crawling nematode is known to follow the resistive force theory with a large anisotropic drag ratio \cite{keaveny_predicting_2017}, which, however, requires a data-fitting process to determine.

As in the synthesized data, we analyze the local tangent angle along the centerline and our shape dimension $N$ is the number of discretized local angles.
The zebrafish sperm flagellar centerline is discretized into $N = 29$ shape dimensions and contains approximately 30 beat periods in $N_t = 300$ time samples, and is characterized by sinusoidal waves with spatially uniform amplitude. The bull sperm data also has $N = 25$ shape dimensions, but is sparser in time, containing approximately 6 beat periods in $N_t = 50$ timepoints, and has larger waveform amplitudes in the distal region. The nematode data has $N = 13$ shape dimensions and 8 beat cycles in $N_t = 515$ time samples, and exhibits a distinct crawling motion with larger fluctuations compared to the flagellar data. For all three qualitatively different datasets, the KAE successfully extracts the limit cycles embedded in the waveforms, as illustrated by the unit-circular latent space dynamics shown in Fig.~\ref{fig:data}(b).

\begin{figure*}[!t]
\centering
    \begin{overpic}
    [width=0.99\linewidth]{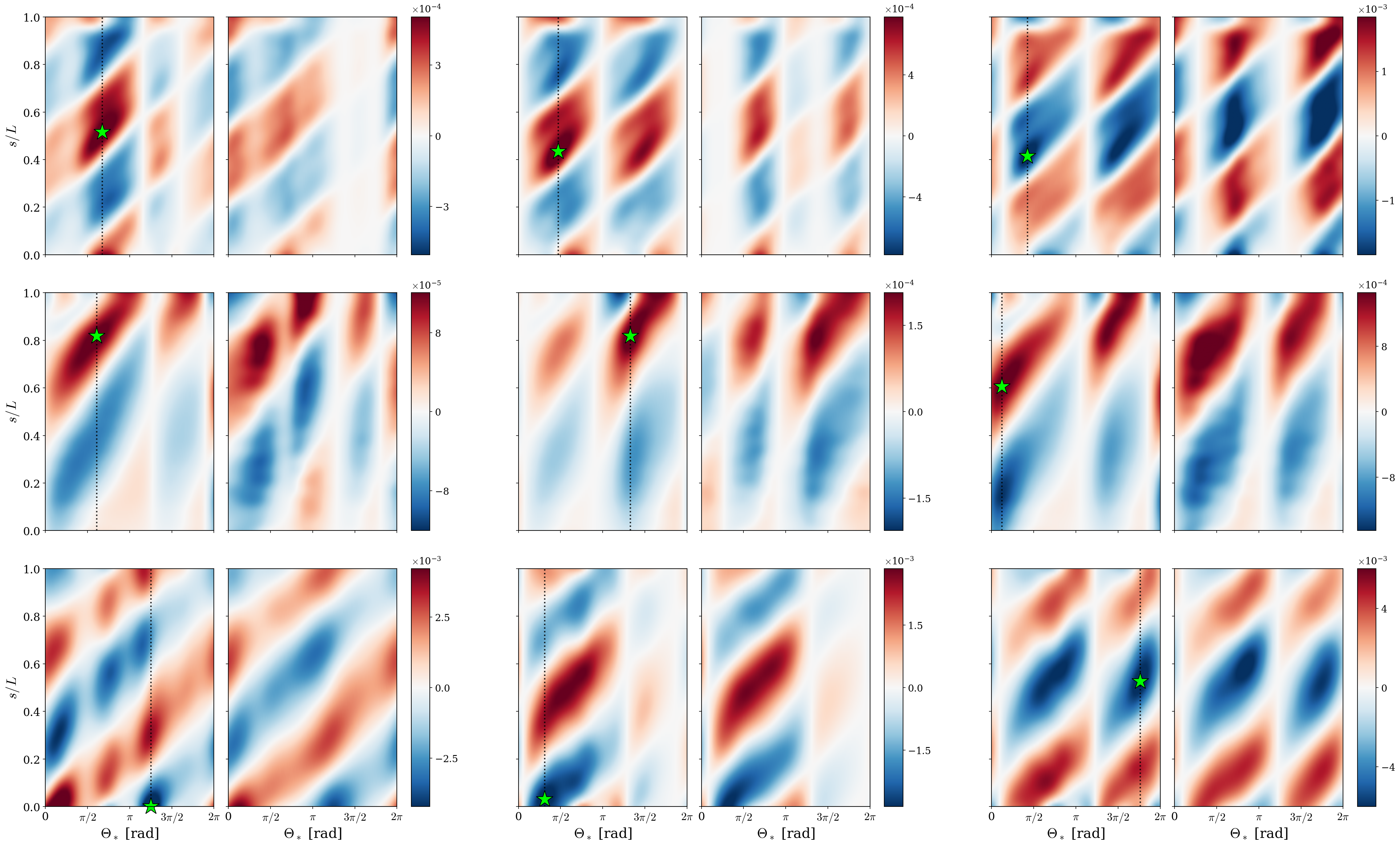}
    \put(-2,60){(a)}
    \put(33,60){(b)}
    \put(67,60){(c)}
    \put(12,62){$Z^x_W(s, \Theta_\ast)/L$}
    \put(44,62){$Z^y_W(s, \Theta_\ast)/L$}
    \put(79,62){$Z^\theta_W(s, \Theta_\ast)$} 
    \put(5,60){data-driven ~~~~~~~~ RFT}
    \put(39,60){data-driven ~~~~~~~~~ RFT}
    \put(73,60){data-driven ~~~~~~~~ RFT} 
    \put(-1,50){\makebox(0,0){\rotatebox{90}{zebrafish sperm}}}
    \put(-1,30){\makebox(0,0){\rotatebox{90}{bull sperm}}}
    \put(-1,10){\makebox(0,0){\rotatebox{90}{\it C. elegans}}}
    \end{overpic} 
\caption{Geometric PSF (a) $Z^x_W(s, \Theta_\ast)/L$, (b) $Z^y_W(s, \Theta_\ast)/L$, and (c) $Z^\theta_W(s, \Theta_\ast)$ for zebrafish sperm, bull sperm, and nematode data, computed by the data-driven method (left) and RFT calculations (right). The maximal absolute value in the data-driven result is indicated by a star symbol in each panel.}
\label{fig:Wcrosscheck}
\end{figure*}

\subsection{Extracting geometric phase from experimental biological data}

Now we extract the averaged trajectory as geometric phase on a stable limit cycle only from the shape and trajectory data, without assuming any mechanical laws. The limit cycle is extracted as the unit circle in latent space via KAE. 
Hence, at each time step $t_n$, the locomotor's trajectory data is labeled by the oscillator phase $\Theta(t_n)$. 
On the limit cycle, the gauge field $A(\Theta)$ is 
expanded in a Fourier series as
\begin{equation}
A(\Theta)=\sum_{k=0}^\infty \tilde{A}_{k}e^{ik\Theta},
\label{eq:gauge_fit}
\end{equation}
which corresponds to a Koopman mode decomposition~\cite{mauroy_use_2012} of the gauge field as an observable. We truncate the series to $K$ modes, with $K=2, 3, 4$ for the zebrafish sperm, bull sperm, and nematode waveform, respectively, chosen to minimize the Bayesian information criterion (BIC)~\cite{egan_automatically_2024}. 

Biological trajectory data $g(t)\in {\rm SE(3)}$ is shown as shaded curves in Fig.~\ref{fig:data}(c), translated and rotated so that the trajectory starts at the origin oriented in the negative $x$ direction at the beginning of each cycle, i.e. whenever the oscillator phase $\Theta(t)\equiv0$ (mod $2\pi$). The zebrafish and bull sperm data are sparse in time (8-10 samples per period), while the nematode data vigorously fluctuates. Despite these sparse and noisy samples, we successfully extract the locomotor's trajectories, which are overlaid as smooth red curves. Further methodological validations are discussed in App.~\ref{app:valid}.

\begin{figure*}[!tb]
    \centering
    \begin{overpic}
    [width=0.9\linewidth]{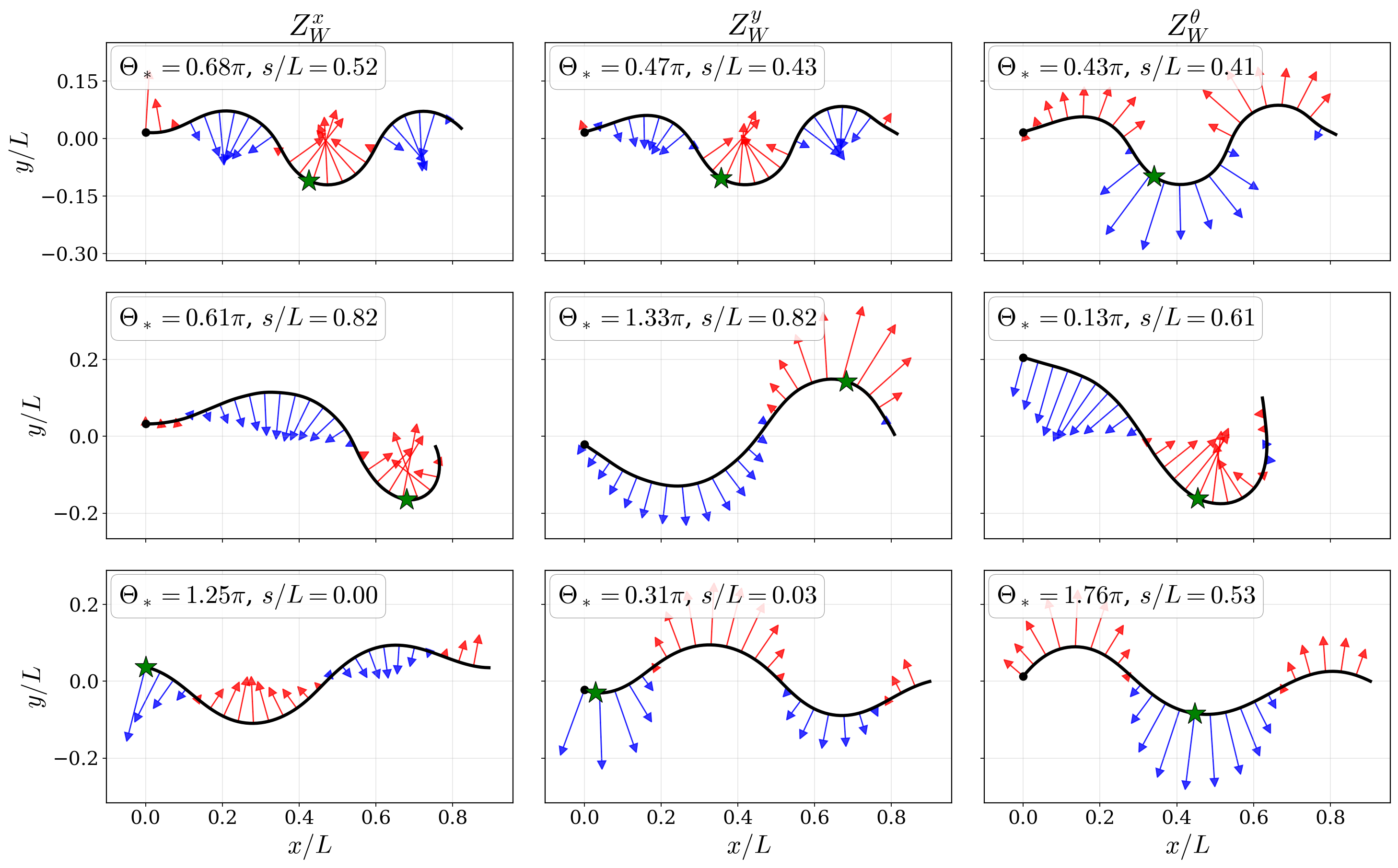}
    \put(9,45){{\scriptsize zebrafish sperm}}
    \put(9,25){{\scriptsize bull sperm}}
    \put(9,7){{\scriptsize {\it C. elegans}}    }
    \put(5,60){(a)}
    \put(36,60){(b)}
    \put(68,60){(c)}
    \end{overpic} 
    \caption{ Sensitivity arrows showing the local normal vector with length proportional to the geometric PSF for the zebrafish sperm (top), bull sperm (middle) and {\it C.elegans} (bottom) shape at the phase of
maximum (a) $|Z^x_W|$, (b) $|Z^y_W|$, and (c) $|Z^\theta_W|$. Green stars mark the position 
of maximum $|Z^\alpha_w|$~($\alpha=x,y,\theta$) and positive/negative values of $Z^\alpha_W$ are shown in red/blue.}
    \label{fig:W}
\end{figure*}

\subsection{Geometric phase sensitivity function}
\label{sec:gpsf}

The phase sensitivity function  $Z_1$ of an oscillator, which we hereafter call the {\it oscillator PSF}, is defined by
\begin{equation}
    \delta \Theta_\ast=\int_0^L Z_1(s,\Theta_\ast)\delta q(s)\,ds.
    \label{eq:OPSF}
\end{equation}
$Z_1(s,\Theta_\ast)$ characterizes the shift in oscillator phase upon a perturbation at the arclength $s$ and the phase $\Theta_\ast$. This phase shift then perturbs the locomotor's position and orientation via $\delta g(\Theta_\ast)=A(\Theta_\ast)\delta \Theta_\ast$, yielding the shift in the geometric phase $W_C$. Writing the motion between $\Theta=0$ and $\Theta=\Theta_\ast$ as 
\begin{equation}
    g_\ast=g^{-1}(0)g(\Theta_\ast),
\end{equation}
the shift in the geometric phase is written as
\begin{equation}
    \delta W_C=g_\ast\exp\left[ A(\Theta_\ast)\delta\Theta_\ast\right]g_\ast^{-1}W_C-W_C,
\end{equation}
which becomes, at leading order in $\delta \Theta_\ast$, 
\begin{equation}
    \delta W_C=Z_2(\Theta_0) W_C\delta \Theta_\ast+O(\delta \Theta_\ast^2), 
\end{equation}
with $Z_2(\Theta_\ast)=g_\ast A(\Theta_\ast) g_\ast^{-1}$. 
This naturally motivates the definition of a {\it geometric phase sensitivity function} (geometric PSF) as 
\begin{equation}
Z_W:=Z_2(\Theta_\ast) Z_1(s, \Theta_\ast),    
\end{equation}
which characterizes the geometric phase shift upon a perturbation of the shape variable via the relation
\begin{equation}
    \delta W_C=\int_0^L Z_W(s,\Theta_\ast)\delta q(s)\,ds+O(\delta q^2)
    \label{eq:GPSF}.
\end{equation}
The geometric PSF for planar locomotion possesses three components $(Z^x_W, Z^y_W, Z^\theta_W)$, which represent the progressive and lateral displacements and the rotational angle upon a perturbation of the local tangent angle. Biologically, a large geometric PSF is beneficial for locomotion and maneuvering, as the locomotor gains a large change of motion from a small shape change.

With the data-driven gauge field in Eq.~\ref{eq:gauge_fit} [Fig.~\ref{fig:data}(c)], we numerically compute the geometric PSF and show in Fig.~\ref{fig:Wcrosscheck} (data-driven) the kymograph of each component, $Z^\alpha_W(s, \Theta_\ast)$ ($\alpha=x,y,\theta$), for the three different biological datasets. 
Stripe patterns in the figure reflect the traveling waveform.  Vertical white stripes in the panels correspond to the phase with $Z_2(\Theta_\ast)=0$. Further discussions on the robustness of the learning process and data-driven calculations of geometric PSF are provided in App.~\ref{app:robustness}

In each component, the maximum of $|Z^\alpha_W(s, \Theta_\ast)|$ is marked by the star symbol and the 
phase $\Theta_\ast$ of maximum $Z^\alpha_W(s, \Theta_\ast)$ is indicated by dotted line. To physically interpret the obtained geometric PSF, we then 
show in Fig.~\ref{fig:W}(a-c) the shape at the phase of maximum $|Z_W^\alpha|$ for each component. The arrow indicates the normal direction to the shape, and the magnitude reflects the size of $Z^\alpha_W$.

In the zebrafish sperm, the geometric PSF is approximately uniform along the flagellum, with a maximum at the middle of the flagellum. In contrast, the bull sperm has its maximum geometric PSF in the distal region, because 
the flagellar amplitude is large here. This larger distal undulation of flagella is a characteristic of mammalian sperm and it has been reported that the distal region  contributes most to propulsion \cite{neal_doing_2020}.  In the nematode case, the maxima of the geometric PSF are located at the head end for both the progressive and lateral translation. This reflects the much smaller tangential drag in crawling on agar relative to normal drag \cite{keaveny_predicting_2017}, resulting in body motions that tend to follow the head trajectory \cite{ishimoto_robust_2025}. These physical interpretations are consistent with the geometric PSF obtained purely from the data.

As seen in Fig.~\ref{fig:W}, the maximum of $|Z^\alpha_W|$ corresponds to a large curvature region in the waveform. As the phase $\Theta_\ast$ increases, the regions with   larger values of the geometric PSF travel along the locomotor's body, which can also be seen as the traveling wave structure in the kymograph of Fig.~\ref{fig:Wcrosscheck}.
\\

\subsection{Comparison with RFT reconstruction}
To further validate our data-driven method, we construct the geometric PSF by assuming mechanical relations of resistive force theory with an anisotropic ratio $\gamma$. Here, we treat $\gamma$ as a free parameter fitted to the biological raw data. After non-dimensionalizing the raw data and normalizing each component of the swimming velocity with its variance, we numerically compute the best-fit anisotropic ratio $\gamma$ by the least squares method. 

In principle, to reconstruct the instantaneous velocity from the resistive force theory, we require additional assumptions for the cell morphology. Here, we assumed that the zebrafish sperm possesses a spherical head of radius $b/L=0.036$ and obtained the best fit anisotropic drag ratio as $\gamma\approx1.62$. For the bull sperm case, we followed Ref.~\cite{friedrich_high-precision_2010} and assumed a spheroidal head of aspect ratio 2, with a major axis of $b/L=0.083$, and we obtained $\gamma\approx1.73$. In the {\it C. elegans} data, we assumed a uniform slender structure and the best-fit value of $\gamma$ is calculated as $\gamma\approx 62.5$. The estimated values of $\gamma$ in the three data are consistent with previous studies \cite{friedrich_high-precision_2010, keaveny_predicting_2017}.

With the estimated anisotropic drag ratio in each data set, we compute the gauge field $A(\Theta)$ via resistive force theory, instead of using Eq.~\ref{eq:gauge_fit}, to obtain the geometric PSF, which is plotted  in Fig.~\ref{fig:Wcrosscheck} (RFT).

Fig. ~\ref{fig:Wcrosscheck} provides a side-by-side comparison of the geometric PSF between the fully data-driven method and reconstruction by the empirical resistive force theory, showing good qualitative agreement,
although the peak magnitudes and positions do not always coincide.
We anticipate that the mismatch reflects the limitations of the RFT reconstruction, because resistive force theory 
only provides the lowest-order asymptotics and neglects non-local hydrodynamics, as well as requiring additional morphological assumptions. 

In contrast, our data-driven approach does not rely on  morphological details or empirical mechanical laws to compute the geometric PSF. The comparison with the RFT reconstruction provides an independent verification of our data-driven methods.

\section{Concluding remarks}
\label{sec:conc}

We have presented a data-driven approach to extract geometric phase in biological locomotion, enforcing only gauge-theoretic structure on the data set without assuming any mechanical laws. Crucially, our approach combines the broad applicability of data-driven methods with a solid foundation in Koopman operator theory. Further, we introduced a geometric PSF that encodes the underlying mechanics directly from the sparse and noisy observations characteristic of biological measurements. 

While only two-dimensional motion was presented in the current study, our data-driven method readily extends to three-dimensional deformation and locomotion \cite{hernandez-herrera_3dt_2026}. 
Since the Koopman phase reduction is applicable to a stable limit torus \cite{mezic_spectrum_2020}, our method is also potentially extendable to more complex dynamics such as the multiple flagella of {\it Chlamydomonas} \cite{wan_lag_2014} and multi-legged locomotion \cite{chong_self-propulsion_2023}, 
although a higher dimensional latent manifold may require a larger amount of data for reconstruction.

Biological locomotion often contains multiple gaits and switches from one to another as seen in forward and backward locomotion of {\it C. elegans} \cite{ishimoto_robust_2025}. Since the current method relies on the existence of a stable limit cycle, it is not applicable to such transient behaviors, although we can extract periodic dynamics from a short time window of repetitive motion.  

In the current study, we analyzed a representative dataset in each species to demonstrate the applicability of our method. Applied across more datasets, our method could be helpful in analyzing the variations within individuals, samples and species \cite{walker_computer-assisted_2020}.  

Whereas data-driven approaches to learning gauge fields have been actively studied in robotic locomotion \cite{deng_data-driven_2024, hu_learning_2025, yang_geometric_2025}, where clean and large data sets are easily available, our methods successfully operate within the severe constraints of biological data. More generally, our study has broad implications beyond biological locomotion, demonstrating how geometric structure can be exploited to infer physical structure from limited and imperfect observations when detailed governing laws are inaccessible.

\section*{Acknowledgements}
P.~H.~H. and K.~I. acknowledge the Japan Society for the Promotion of Science (JSPS) KAKENHI for JSPS International Research Fellow (Grant No. 25KF0222).
K.~I. acknowledges JSPS KAKENHI (Grant Nos. 21H05309 and 24K21517) and the Japan Science and Technology Agency (JST), FOREST (Grant No. JPMJFR212N) and CREST (Grant No. JPMJCR25Q1). 

\section*{Data availability statement}
The code and data needed to reproduce all figures are publicly available at \url{https://github.com/kentaishimoto/kae_biolocomotion.git}.

\appendix
\if
\section{KAE architecture and training details}
\label{app:learning}

In this section, we provide further details of our KAE architecture and learning procedures. 

We implemented encoder and decoder functions using a deep neural network with two hidden layers. The number of nodes is set to $N_1=16$ and $N_2=32$ for learning the synthesized data and the bull sperm data. For the longer time series in the zebrafish sperm and nematode data, we set $(N_1, N_2)=(16, 64)$. The overall number of nodes of the neural network is set, from encoder to decoder, as $N\to N_2\to N_1\to 2\to 2\to N_1\to N_2\to N$, with internal layers with 2 nodes for the latent dynamics  [See also Fig.~\ref{fig:schem}]. We used the $\tanh$ function as an activation function and implemented these deep neural computations via PyTorch.

For training, we used the Adam optimizer with a learning rate of $10^{-3}$ and a weight decay of $10^{-5}$, and the regularization parameters in the loss function were set to $\lambda_{\rm lin}=5.5$ and $\lambda_{\rm rad}=0.2$ in most cases, except for the zebrafish sperm data, for which we used $\lambda_{\rm rad}=2.5$.
The initial value of $\omega$ was guessed from the data set and fixed for the first 40 epochs. In the current data sets, 350 total epochs were sufficient for learning convergence.

 \section{Further analysis on synthesized waveforms}
 \label{app:valid}
\subsection{Stuart-Landau oscillator and oscillator PSF}

The synthesized data is generated by a stochastic oscillator in Eq.~\ref{eq:synth}, whose deterministic part is known as the Stuart-Landau oscillator. Its principal Koopman eigenfunction is analytically obtained as
\begin{equation}
    \psi(r,\phi)=\frac{1}{r}\exp\left( i\Theta(r,\phi)\right),
\end{equation}
with eigenvalue $\mu=-b+i\omega_{\rm eff}$.  Here $\omega_{\rm eff}=\omega+b$ presents the phase velocity at the limit cycle $r=1$, and the function
\begin{equation}
    \Theta(r,\phi)=\phi+\frac{b}{c}\ln r
    \label{eq:isochron}
\end{equation}
is the isochron phase, which properly defines the phase of an oscillator even outside of the limit cycle \cite{nakao_phase_2016}.
Oscillator PSF, $Z_1(s,\Theta)$ introduced in Eq.~\eqref{eq:OPSF} is analytically calculated for the Stuart-Landau oscillator. On the limit cycle, in particular, the oscillator PSF from the deterministic part is written, after normalization, as 
\begin{equation}
    Z_1(s,\Theta)=\frac{\partial q}{\partial \Theta}(s,\Theta)\bigg/\int_0^L\bigg|\frac{\partial q}{\partial \Theta}(s,\Theta)\bigg|^2\,ds.
\end{equation}
The oscillator PSF is also obtained from the learned decoder, with $\bm{q}=\bm{\psi}(\bm{z})$, as
\begin{equation}
   Z_{\rm KAE}=\frac{\partial \bm{q}}{\partial \Theta}={\bf J}\frac{d\bm{z}}{d\Theta},
\end{equation}
where ${\bf J}$ is the Jacobian of the decoder and numerically computed via automated differentiation. Comparisons between analytical representation and the obtained KAE computation, plotted in Fig.~\ref{fig:synth_psf}, show excellent agreement and validate the KAE computation. We also plot the oscillator PSF $Z_{\rm PCA}$ from two-dimensional PCA, which fails to capture the higher frequency modes.

 \begin{figure}[!tb]
     \centering
     \includegraphics[width=0.49\textwidth]{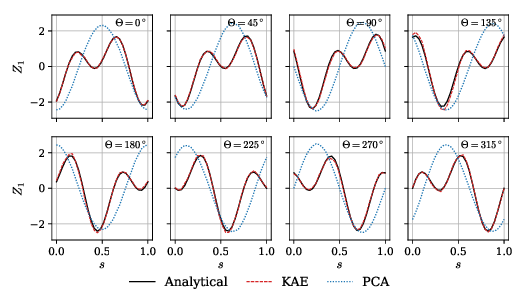}
     \caption{Comparisons of calculated phase response function. The parameters are the same as in Fig.~\ref{fig:synth_mfd}.}
     \label{fig:synth_psf}
 \end{figure}

\subsection{Extracting geometric phase from synthetic trajectory data}

We then extract the geometric phase from fluctuating swimming trajectories. To synthesize the trajectory data, we apply resistive force theory to the shape gait generated by the stochastic dynamics of Eq.~\eqref{eq:synth}. 
 Here we assume that at each time $t$, the instantaneous swimming velocity is determined by the current shape and the deterministic shape change rate, which is calculated from Eq.~\eqref{eq:synth} as
 \begin{equation}
     \dot{q}_{\rm det}(s,t)=\dot{r}_{\rm det}\left(\frac{\partial q}{\partial r}\right)+\dot{\phi}_{\rm det} \left(\frac{\partial q}{\partial \phi}\right).
 \end{equation}

The stochastic increments $dW_r$ and $dW_\phi$  contribute to the evolution of $(r,\phi)$ across time steps but do not enter the instantaneous shape velocity seen by the fluid, considering that the swimmer's internal forcing (e.g., molecular motors) produces deterministic shape changes. This yields a velocity signal $g^{-1}\dot{g} \in \mathfrak{se}(3)$ that is smooth at each instant but modulated by the fluctuating amplitude $r(t)$.

\begin{figure}[!tb]
    \centering
    \begin{overpic}
    [width=0.99\linewidth]{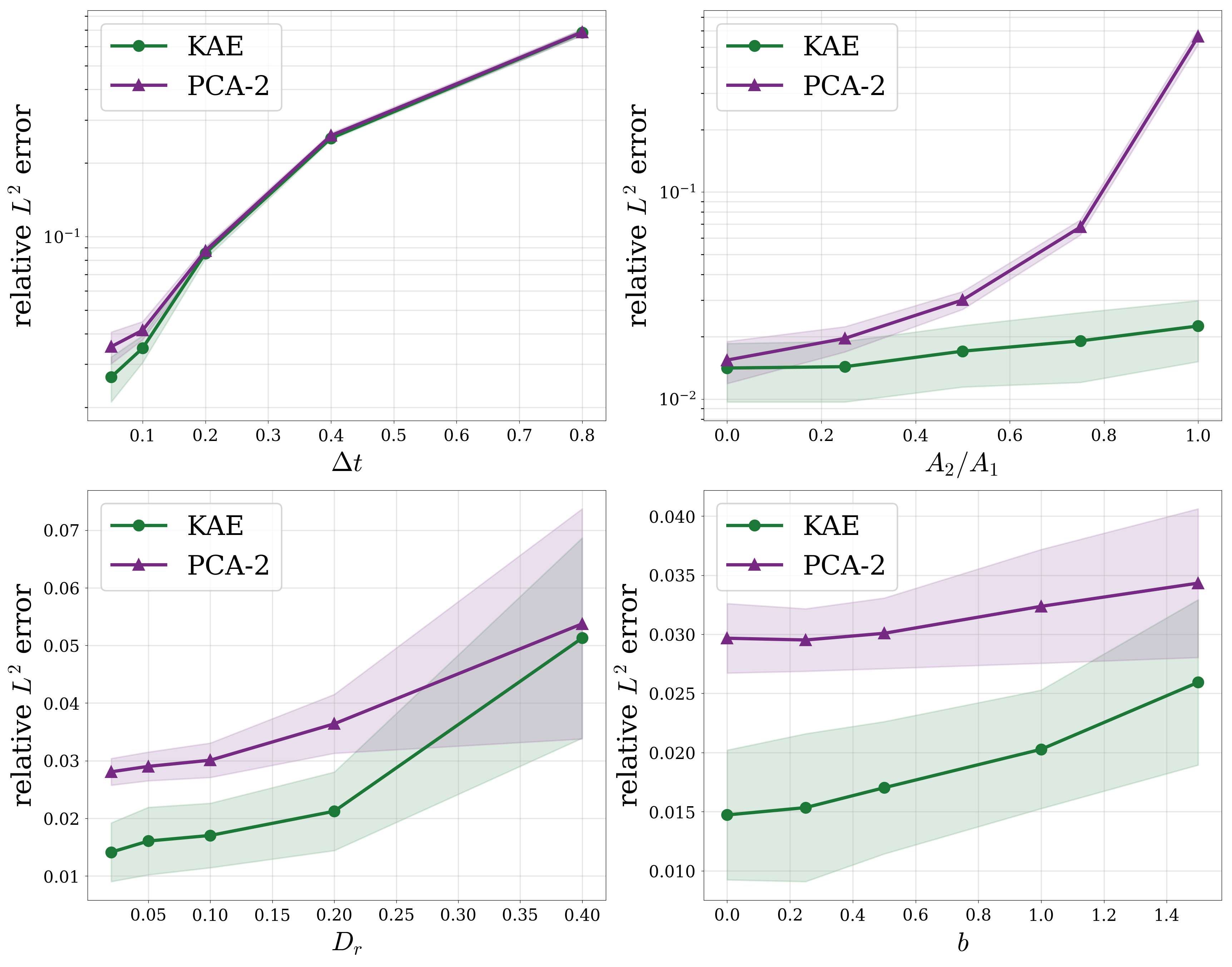}
    \put(0,78){(a)}
    \put(50,78){(b)}
    \put(0,39){(c)}
    \put(50,39){(d)}
    \end{overpic} 
    \caption{Parameter dependence of the trajectory averaging for the synthetic waveform. (a) Sampling time, (b) ratio of second-harmonics, (c) strength of radial diffusion, and (d) nonuniformity of phase velocity. The relative error is plotted with error bars indicating standard deviation over different sequences of stochastic waveforms.}
    \label{fig:ave_benchmark}
\end{figure}

To validate the KAE-based extraction of geometric phase, we compared different methods and took averages using the analytic  isochron phase [Eq.~\ref{eq:isochron}]. Denoting the linear and angular velocity components of the gauge field by the hat symbol as $\hat{A}(\phi)=(A_x, A_y, A_\theta)$, we calculate the $L^2$ relative error defined as
\begin{equation}
    \epsilon=||\hat{A}-\hat{A}_{\rm ref}||/||\hat{A}_{\rm ref}||,
\end{equation}
where $A_{\rm ref}$ is the gauge field estimated by averaging with the analytic expression of the isochron function [Eq.~\ref{eq:isochron}].
We varied the values of sampling $\Delta t$, the ratio $A_2/A_1$ of the second to the first harmonic, radial diffusion $D_r$, and nonuniform phase speed $b$. We tested 20 random sequences of the synthesized waveform and summarize our results in Fig.~\ref{fig:ave_benchmark}, with error bars representing standard deviation. 

We confirm that our KAE-based extraction agrees well with the reference values, with relative errors of $\epsilon =O(10^{-2})$ for the parameter set used in Fig.~\ref{fig:synth_mfd}. We also compare results from different methods. A protophase may be obtained as the angle in the two-dimensional PCA plane. Although the averaging with this protophase performs similarly well to KAE-based averaging for this parameter set [see the curve labeled `PCA-2' in Fig.~\ref{fig:ave_benchmark}(a)], the error of PCA-based averaging increases substantially with the second harmonics ratio [Fig.~\ref{fig:ave_benchmark}(b)].

In the panel with different $D_r$  [Fig.~\ref{fig:ave_benchmark}(c)], the KAE-based averaging shows a similar relative error to the PCA result in the upper end of the range of $D_r$ considered. However, this has no theoretical basis in the PCA algorithm and is likely coincidental. Panel (d) examines the impact of the parameter $b$ for the nonuniform phase velocity, showing only minor effects on the relative error.

\section{Data acquisition}
 \label{app:dataacq}

Sperm flagellum centerlines in the lab frame were obtained from processed centerline data provided in Ref.~\cite{guasto_flagellar_2020}. We used the single available bull sperm dataset, and file number 6, run number 3 from the zebrafish datasets. Nematode centerlines in the lab frame were extracted by thresholding each frame of Supplementary Movie 3 of Ref.~\cite{pierce-shimomura_genetic_2008}, skeletonizing the worm body, and resampling each centerline to 14 equally spaced points along arclength. The body-fixed frame tangent angles were obtained by computing a global rotation rate $\Omega$, assumed constant in time, which minimizes 
\begin{equation}
    \text{Var}_k\left[{\frac{1}{N}\sum_{j = 1}^N q^\text{(lab)}(s_j, t_k)-\Omega t_k}\right],
\end{equation} where $q^\text{(lab)}(s_j, t_k)$ denotes the lab-frame tangent angle at arclength $s_j$ and time $t_k$.

\section{Robustness of the learning process in  real data analysis}
\label{app:robustness}
To examine the robustness of the learning process and the calculations of geometric PSF in Sec.~\ref{sec:expr}, we tested the sensitivity by doubling and halving the network sizes, $N_1$ and $N_2$, and the regularization coefficients, $\lambda_{\rm lin}$ and $\lambda_{\rm rad}$.

We confirmed for zebrafish sperm and bull sperm data that the KAE successfully learns the limit cycle with unit radius and isochron phase, regardless of changes in the network size and regularization parameters. 
The robustness of the obtained geometric PSF is measured by the $L^2$ relative error defined as
\begin{equation}
    \epsilon_\alpha=||Z^\alpha_W-Z^\alpha_{W, {\rm ref}}||/||Z^\alpha_{W, {\rm ref}}||,
\end{equation}
where $\alpha=\{x,y, \theta\}$ and $Z^\alpha_{W, {\rm ref}}$ is the data shown in Fig.~\ref{fig:Wcrosscheck}. In the zebrafish sperm and bull sperm cases, the relative error was confirmed to be sufficiently small ($\epsilon_\alpha\leq 0.02$). The position of the peak of $|Z^\alpha_W|$ and the physical interpretations discussed in Sec.~\ref{sec:gpsf} are also confined to be robust under changes in learning processes.

In the {\it C. elegans} data, our KAE learning sometimes failed to extract the limit cycle in the latent space for particular changes in the network size and parameters. 
Noticeable declines in success rate were found 
when the network sized is doubled, when $\lambda_{\rm lin}$ is halved, and when $\lambda_{\rm rad}$ is doubled. The choice of network size should depend on the data size, and we do not need a large network as our latent manifold is a simple unit circle. The size of regularization parameters should be properly chosen to prevent the learned manifold from shrinking to a point, which is a trivial mapping to construct the linear dynamics in latent space.

Although the choice of network size and regularization parameters are common practical issues in machine learning architectures, in the current problem, failed learning was easily detected by checking the radius of the learned manifold and the value of the learned phase velocity, $\omega$. 
More importantly, once the KAE passes this intuitive test, the obtained geometric PSF is found to be robust under changes in learning parameters. We confirmed that the relative $L^2$ error of the geometric PSF was $\epsilon_\alpha\leq 0.07$ and that the physical arguments along Figs.~\ref{fig:Wcrosscheck} and \ref{fig:W} are also robust.



\bibliographystyle{apsrev4-2}
\bibliography{refs}

\end{document}